\documentclass{JUSTC}

\usepackage{algorithm,algorithmicx,algpseudocode}
\usepackage{graphicx,times}
\usepackage{amssymb,amsmath}
\usepackage{threeparttablex}
\usepackage{longtable}
\usepackage{booktabs}
\usepackage[pagebackref=true]{hyperref}

\newcommand{\upcite}[1]{\textsuperscript{\textsuperscript{\cite{#1}}}}

\type{Manuscript type}
\title{The \textit{NuSTAR} Extragalactic Surveys: Source Catalogs from the Extended \textit{Chandra} Deep Field-South and the \textit{Chandra} Deep Field-North}

\author[1,2,\Letter]{Tianyi Zhang}
\author[1,2,\Letter]{Yongquan Xue}
\email{zty123@mail.ustc.edu.cn}
\email{xuey@ustc.edu.cn}

\runningtitle{The \textit{NuSTAR} E-CDF-S and CDF-N Surveys}
\runningauthor{Zhang \& Xue}

\address{CAS Key Laboratory for Research in Galaxies and Cosmology, Department of Astronomy, University of Science and Technology of China, Hefei 230026, China}
\address{School of Astronomy and Space Sciences, University of Science and Technology of China, Hefei 230026, China}

\GraphicalAbstract{
\begin{figure}
\centering
\includegraphics[width=\textwidth]{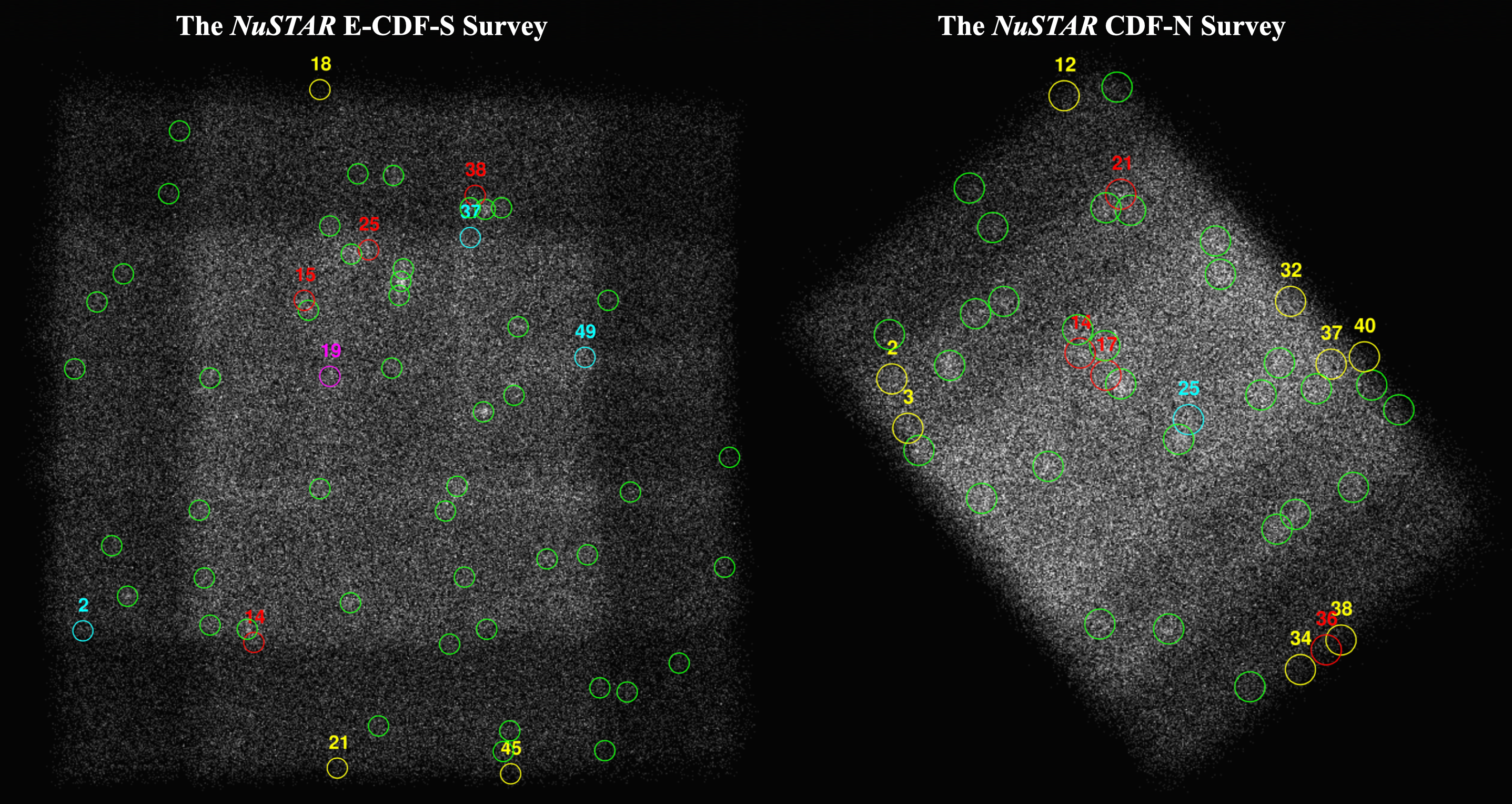}
	\caption*{Stacked \textit{NuSTAR} E-CDF-S and CDF-N science mosaics in the 3-24 keV band.}
\label{fig:graph_abs}
\end{figure}
}

\abstract{
	We present a routinized and reliable method to obtain source catalogs from the \textit{Nuclear Spectroscopic Telescope Array} (\textit{NuSTAR}) extragalactic surveys of the Extended \textit{Chandra} Deep Field-South (E-CDF-S) and \textit{Chandra} Deep Field-North (CDF-N). The \textit{NuSTAR} E-CDF-S survey covers a sky area of $\approx30'\times30'$ to a maximum depth of $\sim$ 230 ks corrected for vignetting in the 3--24 keV band, with a total of 58 sources detected in our E-CDF-S catalog; the \textit{NuSTAR} CDF-N survey covers a sky area of $\approx7'\times10'$ to a maximum depth of $\sim$ 440 ks corrected for vignetting in the 3--24 keV band, with a total of 42 sources detected in our CDF-N catalog that is produced for the first time. We verify the reliability of our two catalogs by crossmatching them with the relevant catalogs from the \textit{Chandra} X-ray observatory, and find that the fluxes of our \textit{NuSTAR} sources are generally consistent with that of their \textit{Chandra} counterparts. Our two catalogs are produced following the exactly same method and made publicly available, thereby providing a uniform platform that facilitates further studies involving these two fields. Our source-detection method provides a systematic approach for source cataloging in other \textit{NuSTAR} extragalactic surveys.}


\begin{document}

\maketitle

\section{Introduction}           %% first-level sections will be auto-capitalized
Extragalactic X-ray surveys are efficient at identifying and characterizing highly reliable and fairly complete samples of active galactic nuclei (AGNs), given the following reasons (see, e.g., Brandt and Hasinger\upcite{Brandt2005}; Brandt and Alexander\upcite{Brandt2015}; Xue\upcite{Xue:2017rpk}, for reviews]. Firstly, X-ray emission is nearly a universal feature of luminous AGNs, which can be produced in various accretion disk (plus corona) models for AGNs (e.g., Yuan and Narayan\upcite{YN2014}). Secondly, X-ray emission, especially hard X-ray emission ($\geq10$ keV), can penetrate  through materials with Hydrogen column densities even up to $N_{\rm H}\sim10^{25}\,\mathrm{{cm}^{-2}}$, which is key to excavating the majority of the AGN family, i.e., highly obscured and even Compton-thick AGNs (e.g., Li et al.\upcite{Li2019,Li2020}).
Thirdly, X-ray emission is subject to minimal dilution by host-galaxy stellar emission, being powerful to probe the immediate vicinity of supermassive black holes (SMBHs) in AGNs even at high redshifts.
Lastly, the production of X-ray spectrum goes through numerous line and continuum emission processes, and a high-quality X-ray spectrum is effective to infer physical conditions near the central SMBH.\par
X-ray surveys have resolved a very large portion ($\approx 80$--90\%) of the cosmic X-ray background (CXRB) up to $\approx$10 keV, with AGNs being the dominant contributor (e.g., Hickox and Markevitch\upcite{Hickox2006}; Xue et al.\upcite{Xue2011a}; Xue et al.\upcite{Xue2012}; Lehmer et al.\upcite{Lehmer2012}; Luo et al.\upcite{Luo2017}), but the resolved fraction around the peak of the CXRB at $\approx$ 20--40 keV has been very low ($<10$\%; see, e.g., Brandt and Hasinger\upcite{Brandt2005}; Brandt and Alexander\upcite{Brandt2015}; Harrison et al.\upcite{Harrison2016}). The \textit{Nuclear Spectroscopic Telescope Array} (\textit{NuSTAR}), the
first focusing high-energy X-ray (3--79 keV) telescope in orbit, has largely broadened the window of X-ray observations\upcite{Harrison2013a}. The \textit{NuSTAR} surveys have resolved $\approx$33--39\% of the 8--24 keV CXRB\upcite{Harrison2016}, thereby helping us better understand the contribution of highly obscured and Compton-thick AGNs to the CXRB.\par
The \textit{Chandra} Deep Fields (CDFs), consisting of the \textit{Chandra} Deep Field-South (CDF-S; Luo et al.\upcite{Luo2017}, hereafter L17), \textit{Chandra} Deep Field-North  (CDF-N; Xue et al.\upcite{Xue2016a}, hereafter X16) and extended \textit{Chandra} Deep Field-South (E-CDF-S; X16), are important sky areas for studying, e.g., AGN demography, physics, and evolution\upcite{Xue:2017rpk}. The \textit{Chandra} X-ray observatory has accumulated $\sim$7 Ms exposure in the CDF-S (L17), the deepest X-ray exposure ever made, which provides a large sample of AGNs at $z\approx 0$--5 for powerful statistical studies. As a parallel field to the CDF-S and being the second deepest X-ray survey, the 2~Ms CDF-N (X16) effectively complements the 7~Ms CDF-S, accounting for cosmic variance and enabling comparative studies between fields. \textit{NuSTAR} has also observed the CDFs for complementary studies over 10 keV, and has completed a series of additional extragalactic hard X-ray surveys\upcite{Harrison2016}. Mullaney et al.\upcite{Mullaney2015a} (hereafter M15) has released a source catalog from the \textit{NuSTAR} E-CDF-S survey; however, the \textit{NuSTAR} CDF-N catalog is still absent.\par
In this work, we put forward a uniform and reliable method to process the \textit{NuSTAR} E-CDF-S and CDF-N observations and perform source detection. Referring to the previous \textit{NuSTAR} E-CDF-S cataloging work (M15), we obtain both the \textit{NuSTAR} E-CDF-S and CDF-N source catalogs in a routinized and unifrom way. We describe the production of both catalogs in Section~\ref{sec:E-CDF-S_cat} and Section~\ref{sec:CDF-N_cat}, respectively, where, for brevity, the data reduction and source detection is introduced in detail only for the E-CDF-S. We summarize our results in Section~\ref{sec:summary}. We use J2000.0 coordinates and a cosmology of $H_0=71\,\mathrm{km\,s^{-1}\,Mpc^{-1}}$, $\Omega_{\rm M}=0.27$, and $\Omega_{\Lambda}=0.73$.

\section{Production of the \textit{NuSTAR} E-CDF-S Point-Source Catalog}
\label{sec:E-CDF-S_cat}
\subsection{Data Reduction}
\label{sec:data}
We collect 33 valid observations from the \textit{NuSTAR} E-CDF-S survey that cover a sky area of $\approx30'\times30'$, almost each of which has an effective exposure of $\approx 45$\,ks. The details of these observations are presented in Table~\ref{tab:table1}.

\begin{center}
\begin{ThreePartTable}

\begin{TableNotes}\footnotesize
\item[] \textbf{Notes}: Obs.\,ID\tnote{1} is a unique identification number specifying the \textit{NuSTAR} observation; Obs.\,Name\tnote{2} gives the designation of the target at which \textit{NuSTAR} was pointing; Obs.\,Date\tnote{3} is the start time of the observation; RA\tnote{4} and DEC\tnote{5} give the J2000.0 Right Ascension and the Declination of the \textit{NuSTAR} pointing position; $t_{\rm eff}$\tnote{6} is the effective exposure time (in ks) after background filtering (see Section~\ref{sec:flare}).
\end{TableNotes}

\begin{longtable}{cccccc}
% 首页表头
	\caption{Details of the \textit{NuSTAR} E-CDF-S Observations}	\label{tab:table1} \\
\toprule[1.5pt]
	Obs.\,ID\tnote{1} & Obs.\,Name\tnote{2} & Obs.\,Date\tnote{3} & RA\tnote{4} & DEC\tnote{5} & $t_{\rm eff}$\tnote{6} \\
\midrule[1pt]
\endfirsthead
% 续页表头
	\caption[]{Details of the \textit{NuSTAR} E-CDF-S Observations (continued)} \\
\toprule[1.5pt]
	Obs.\,ID\tnote{1} & Obs.\,Name\tnote{2} & Obs.\,Date\tnote{3} & RA\tnote{4} & DEC\tnote{5} & $t_{\rm eff}$\tnote{6} \\
\midrule[1pt]
\endhead
% 首页表尾
\hline
\multicolumn{6}{r}{\textit{Table~\ref{tab:table1} continued}}
\endfoot
% 续页表尾
\bottomrule[1.5pt]
\insertTableNotes
\endlastfoot
	60022001002 & ECDFS\_MOS001 & 2012 Sep 28 & 52.93 & $-$27.97 & 49.0 \\
	60022002001 & ECDFS\_MOS002 & 2012 Sep 29 & 52.93 & $-$27.97 & 50.3 \\
	60022003001 & ECDFS\_MOS003 & 2012 Sep 30 & 52.93 & $-$27.97 & 50.2 \\
	60022004001 & ECDFS\_MOS004 & 2012 Oct 01 & 52.93 & $-$27.97 & 50.9 \\
	60022005001 & ECDFS\_MOS005 & 2012 Oct 02 & 53.06 & $-$27.86 & 50.5 \\
	60022006001 & ECDFS\_MOS006 & 2012 Oct 04 & 53.06 & $-$27.86 & 49.2 \\
	60022007002 & ECDFS\_MOS007 & 2012 Nov 30 & 53.06 & $-$27.86 & 51.7 \\
	60022008001 & ECDFS\_MOS008 & 2012 Dec 01 & 53.06 & $-$27.86 & 51.7 \\
	60022009001 & ECDFS\_MOS009 & 2012 Dec 03 & 53.18 & $-$27.75 & 50.3 \\
	60022010001 & ECDFS\_MOS010 & 2012 Dec 04 & 53.18 & $-$27.75 & 51.2 \\
	60022011001 & ECDFS\_MOS011 & 2012 Dec 05 & 53.18 & $-$27.75 & 51.7 \\
	60022012001 & ECDFS\_MOS012 & 2012 Dec 06 & 53.18 & $-$27.75 & 52.1 \\
	60022013001 & ECDFS\_MOS013 & 2012 Dec 07 & 53.31 & $-$27.64 & 52.5 \\
	60022014001 & ECDFS\_MOS014 & 2012 Dec 08 & 53.31 & $-$27.64 & 52.9 \\
	60022015001 & ECDFS\_MOS015 & 2012 Dec 09 & 53.31 & $-$27.64 & 53.2 \\
	60022016001 & ECDFS\_MOS016 & 2012 Dec 10 & 53.31 & $-$27.64 & 50.1 \\
	60022016003 & ECDFS\_MOS016 & 2013 Mar 15 & 52.93 & $-$27.64 & 51.7 \\
	60022015003 & ECDFS\_MOS015 & 2013 Mar 17 & 52.93 & $-$27.64 & 51.2 \\
	60022014002 & ECDFS\_MOS014 & 2013 Mar 18 & 52.93 & $-$27.64 & 51.4 \\
	60022013002 & ECDFS\_MOS013 & 2013 Mar 19 & 52.93 & $-$27.64 & 49.7 \\
	60022012002 & ECDFS\_MOS012 & 2013 Mar 20 & 53.06 & $-$27.75 & 49.7 \\
	60022011002 & ECDFS\_MOS011 & 2013 Mar 21 & 53.06 & $-$27.75 & 48.9 \\
	60022010002 & ECDFS\_MOS010 & 2013 Mar 22 & 53.06 & $-$27.75 & 32.5 \\
	60022010004 & ECDFS\_MOS010 & 2013 Mar 23 & 53.06 & $-$27.75 & 16.4 \\
	60022009003 & ECDFS\_MOS009 & 2013 Mar 24 & 53.18 & $-$27.75 & 49.5 \\
	60022008002 & ECDFS\_MOS008 & 2013 Mar 25 & 53.18 & $-$27.86 & 49.8 \\
	60022007003 & ECDFS\_MOS007 & 2013 Mar 26 & 53.18 & $-$27.86 & 49.6 \\
	60022006002 & ECDFS\_MOS006 & 2013 Mar 27 & 53.18 & $-$27.86 & 48.7 \\
	60022005002 & ECDFS\_MOS005 & 2013 Mar 28 & 53.31 & $-$27.86 & 49.3 \\
	60022004002 & ECDFS\_MOS004 & 2013 Mar 29 & 53.31 & $-$27.97 & 48.8 \\
	60022003002 & ECDFS\_MOS003 & 2013 Mar 30 & 53.31 & $-$27.97 & 49.0 \\
	60022002002 & ECDFS\_MOS002 & 2013 Mar 31 & 53.31 & $-$27.97 & 49.0 \\
	60022001003 & ECDFS\_MOS001 & 2013 Apr 01 & 52.93 & $-$27.97 & 48.2 \\
\end{longtable}
\end{ThreePartTable}
\end{center}

\subsubsection{Flaring episodes}
\label{sec:flare}
As \textit{NuSTAR} is composed of two focal plane modules (i.e., FPMA and FPMB), each of the 33 observations results in two event files. We use the program \textit{nupipeline} of the \textit{NuSTAR} data analysis software NuSTARDAS to generate 66 initial event files with default parameters. Following M15, full-field lightcurves in the entire energy band (i.e., 3--78\,keV) with a bin size of 20~s are produced to inspect the influence of flaring events. The \textit{dmgti} tool of the \textit{Chandra} Interactive Analysis of Observations (CIAO) is used to make a user-defined good-time interval (GTI) file to avoid background flaring when the average binned count rate exceeds 1.5 cts\,s$^{-1}$ in the light curves. Taking the GTI files into account, we run \textit{nupipeline} again to obtain the 66 cleaned event files. Following Alexander et al.\upcite{Alexander2013}, the final cleaned event files are split into three standard energy bands, 3--8 (soft band; S), 8--24 (hard band; H), and 3--24 keV (full band; F), respectively.

\subsubsection{Science, exposure, and background mosaics}
\label{sec:mosaic}
From the cleaned event files, we produce exposure maps with the NuSTARDAS program \textit{nuexpomap}. For the effects of vignetting, the same energy correction values as those in M15 are adopted to generate the effective exposure maps, i.e., 5.42, 13.02, and 9.88 keV for the soft, hard, and full bands, respectively.
The E-CDF-S reaches a maximum depth of $\sim$ 230 ks corrected for vignetting in the full band.\par
\begin{figure}
\centering
\includegraphics[width=0.5\textwidth]{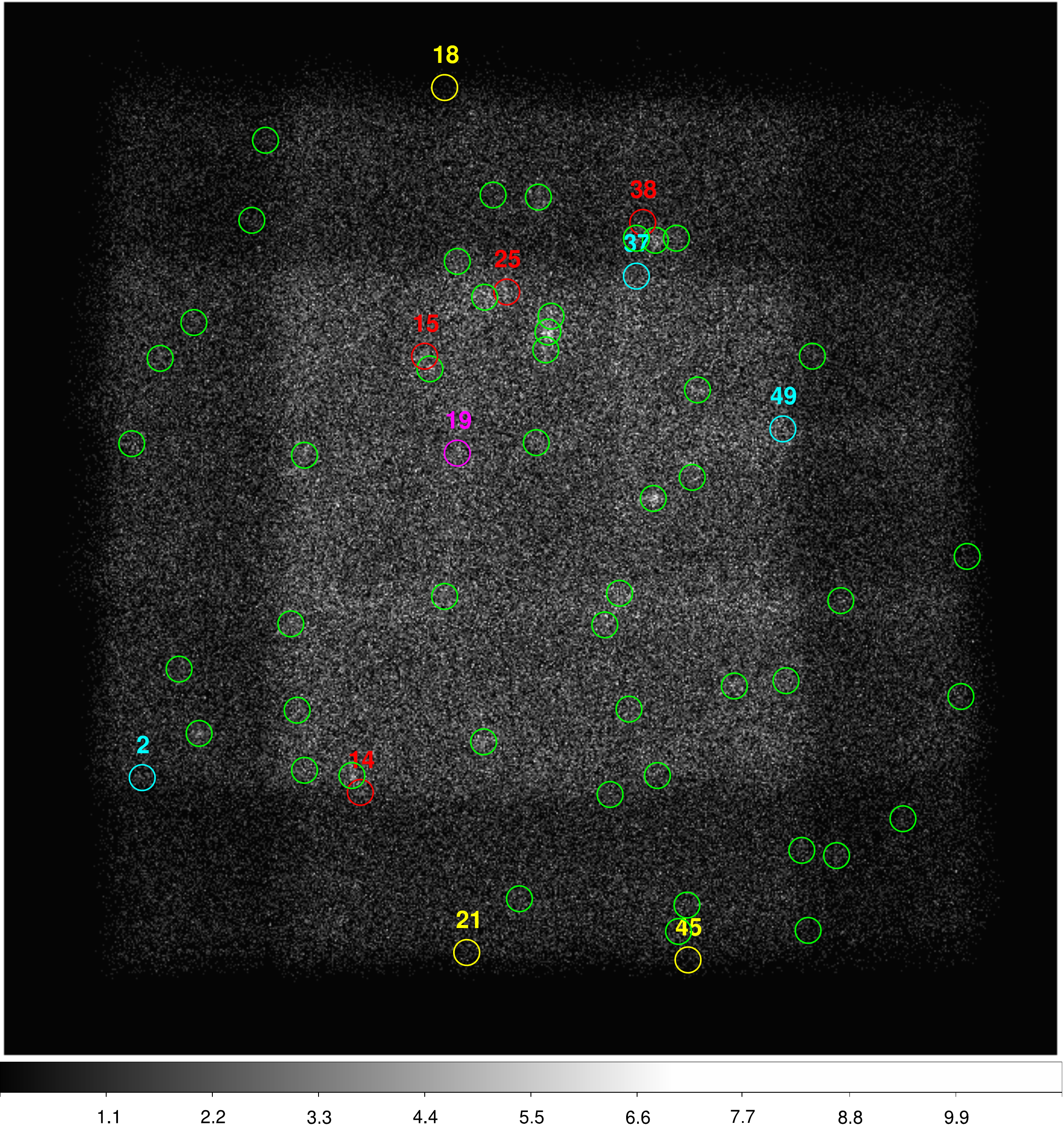}
\caption{Stacked \textit{NuSTAR} E-CDF-S science mosaic in the full band, with a total of 58 sources plotted as circles. The green (47/58) and red (4/58; being less significant detections) sources have the \textit{Chandra} 250\,ks E-CDF-S counterparts within $r_{\rm m}=30''$. The magenta source (1/58) does not have any \textit{Chandra} 250\,ks E-CDF-S counterparts but can be matched to a \textit{Chandra} 7\,Ms CDF-S source.
Among the unmatched sources (6/58), 3 yellow sources reside in the very edges of the mosaic and 3 cyan sources reside in the chip-gap areas. Numbers are source XIDs. The bottom color bar indicates the counts per pixel.}
\label{fig:E-CDF-S_mosaic}
\end{figure}

Due to the high count-rate backgrounds in the \textit{NuSTAR} E-CDF-S observations, we generate model background maps using the IDL software \textit{nuskybgd}\upcite{Wik2014a}. Following the  similar strategy adopted by M15, we choose four large (i.e., radius of $3'$) circular regions centered on the four chips of the detector as our background regions. With the user-defined regions, the \textit{nuskybgd} software can extract and fit the corresponding spectra in XSPEC with the preset models and derive the best-fit parameters. These parameters are used to generate ``fake" background images of the observations. Using the FTOOLS task XIMAGE, these simulated images are collected and merged into background mosaics weighted by the corresponding exposure maps; similarly, using XIMAGE, the stacked science mosaics (see Figure~\ref{fig:E-CDF-S_mosaic}) are directly produced from the cleaned event files.\par

We note that in the newest version of \textit{nuskybgd}, the use of the ``nuabs" XSPEC model has been phased out of \textit{nuskybgd} routines. However, we find that when ``nuabs'' is removed, the model background counts are significantly lower than what they should be. Consequently, we turn to using the old version that includes the ``nuabs'' model in the spectral fitting process.

\subsection{Source Detection}
\label{sec:detect}
As shown in Figure~\ref{fig:E-CDF-S_mosaic}, traditional source detection methods (e.g., WAVDETECT\upcite{Freeman2002} and ACIS Extract\upcite{AE2010} adopted in Xue et al.\upcite{Xue2011a}; Xue et al.\upcite{Xue2016a}; Luo et al.\upcite{Luo2017}) are invalid due to the heavy background. Following the general strategy adopted for \textit{NuSTAR} surveys (M15; Masini et al.\upcite{Masini2018}), we use the incomplete Gamma (igamma) function (see Georgakakis et al.\upcite{Georgakakis2008}) in the \textit{Scipy.special} package to produce false probability ($P_{\rm false}$) maps for source detection:
$$P_{\rm false}={\rm igamma}(N_{\rm Sci}, N_{\rm Bgd}),$$
where $N_{\rm Sci}$ and $N_{\rm Bgd}$ represent the photon counts within one region at the same position in the science and background mosaics, respectively. The $P_{\rm false}$ value gives the probability that a signal with $N_{\rm Sci}$ counts is purely due to random fluctuation given the background of $N_{\rm Bgd}$, which means that the signal is more likely to be real as $P_{\rm false}$ decreases.\par
We smooth the science and background mosaics with top-hat functions of different radii, with the former ($10''$) being smaller for finer structures and the latter ($20''$) being larger to decrease the background influence. The $P_{\rm false}$ maps are produced using three methods: (1) $P_{\rm false}$ value at the position ($x$,$y$) is directly derived by igamma(Sci($x$,$y$), Bgd($x$,$y$)), where the resulting $P_{\rm false}$ maps are called $P_{r0}$ maps; (2) At the position ($x$,$y$), we perform aperture photometry with a circular region of radius $10''$ on the mosaics, then calculate the $P_{\rm false}$ value from igamma(Sci$_{10''}(x,y)$,Bgd$_{10''}(x,y))$, where the resulting $P_{\rm false}$ maps are called $P_{r10}$ maps; (3) The same procedure as method 2 but using a $20''$ radius aperture is adopted to obtain the $P_{r20}$ maps. Considering the potential signals residing in the local minimums of the $P_{\rm false}$ maps, we produce the inverse $P_{\rm false}$ maps using $\log({1/P_{\rm false}})$ to identify peaks with the SExtractor source-detection algorithm\upcite{Bertin1996}.\par
Some modifications on the default SExtractor configuration file \textit{default.sex} are listed below:
\begin{enumerate}
\item
DETECT\_MINAREA is set to 1. Because we smooth the input maps, even one pixel in the 1/$P_{\rm false}$ maps can be considered as a potential source.
\item
THRESH\_TYPE is set to ABSOLUTE. Under this condition, DETECT\_THRESH represents the detection threshold above which a signal in the maps can be considered to be significant.
\item
FILTER and CLEAN are set to N(o). We do not need these processes as what we deal with here are not real observational images.
\end{enumerate}

With proper DETECT\_THRESH values, SExtractor is able to detect potential sources in the three standard bands to produce our initial catalogs.

\subsubsection{Simulations}
We perform a series of simulations to determine DETECT\_THRESH. Because the science mosaics are smoothed with the $10''$-radius circular top-hat function (see Section~\ref{sec:detect}), we split the background images into several square cells with side length of $20''$. For every background image, Poisson realizations are performed on these cells to make up a ``science'' image from pure Poisson fluctuation. The aforementioned procedures are then performed on these Poisson realizations, and SExtractor should detect no sources in the corresponding inverse $P_{\rm false}$ maps with proper thresholds.\par
The simulations are repeated for 100 times, in each of which we produce inverse $P_{\rm false}$ maps with three different aperture radii (i.e., $0''$, $10''$, and $20''$) in the three standard bands. DETECT\_THRESH is set to the value above which SExtractor can find no more than $N$ signals in these simulated maps per aperture radius per band, and $R=N/100$ represents the false detection rate. These thresholds are applied to the real inverse $P_{\rm false}$ maps for source detection.

\subsubsection{Final Catalog Production}
We first set $R=0.1$ for loose thresholds, with which our algorithm is run on the $P_{r0}$, $P_{r10}$, and $P_{r20}$ maps to generate three seed lists in each band. These seed lists are produced from different $P_{\rm false}$ maps, thus there might be multiple nearby detections belonging to one same source. To identify every unique source, we merge the seed lists for each band, then run a modified friends-of-friends algorithm (hereafter FOF) on these merged lists for deduplication.\par
FOF is common in cosmology for identifying groups in dense fields\upcite{Feng2017}, which demands that any particle (or galaxy) that finds another one within a distance $l$ (called linking length) should be linked to it to form a group. Sources in the merged lists are split into group sources and isolated sources by FOF with a linking length of $30''$. For each group, members are ranked by their inverse $P_{r20}$ values (even if being initially detected from other $P_{\rm false}$ maps), then compete against any others within a $30''$ radius, in which a member of higher value can survive and participate in a next competition. After several rounds of competitions, the final survivals from the group sources are combined with the isolated sources to make up the deduplicated seed lists for each band.\par
The deduplicated seed lists from different bands are combined into one seed list and then split by FOF in the same way. For each group in the combined list, members are ranked by their inverse $P_{r20}$ values in the full band (regardless of whether being detected in this band), then participate in competitions against each other. After deduplication, the remaining sources are collected to construct our seed catalog with $R=0.1$. The details are presented in Table~\ref{tab:seed_E-CDF-S}.\par

\begin{table}[th]
\caption{The \textit{NuSTAR} E-CDF-S Cataloging Process}
\label{tab:seed_E-CDF-S}
\centering
\resizebox{\textwidth}{!}{
\begin{tabular}{cccccc}\hline\hline
	Seed Catalog		& Source \# 	& Source \# 	& Source \# 	& Source \#	& Source \# \\
	($R=0.1$)		& ($P_{r0})$	& ($P_{r10}$)	& ($P_{r20}$)	& (Total)	& (After Deduplication) \\
	\hline
	3--8 keV 	& 33		& 35		& 35		& 103	& 46		\\
	8--24 keV 	& 17		& 13		& 20		& 50	& 23		\\
	3--24 keV	& 42		& 41		& 55		& 138	& 66		\\
	\hline\hline
	Final Catalog	& Source \# 	& Source \# 	& Source \# 	& Source \# 	& Source \# \\
	  ($R=0.01$ \& $P_{r20}$)     	& (3--8 keV)	& (8--24 keV)	& (3--24 keV)	& (3--8 \& 8--24 keV)	& (Total) \\
	\hline
	Before Deblending	& 33	& 13		& 54		& 9		& 58		\\
	After Deblending & 33	& 13		& 50		& 9		& 54		\\
	\hline
\end{tabular}
}
\end{table}

To produce a reliable final catalog, we then set $R=0.01$, which means only 1 false signal being detected in 100 simulations. The thresholds of the inverse $P_{r20}$ maps are 3.99, 4.36 and 4.33 in the soft, hard, and full bands, respectively, corresponding to $\approx99\%$ reliability\upcite{Civano2015a}. We remove the sources that do not meet any of our final thresholds and construct the final catalog with the remaining ones.\par
The E-CDF-S final catalog contains 58 sources, each of which is detected in at least one of the three standard bands. Of these 58 sources, 33, 13, and 54 are detected in the soft, hard, and full bands, respectively; 3, 1, 21 are detected only in the soft, hard, and full bands, respectively; no source is detected in exactly the soft and hard bands, 21 in exactly the soft and full bands, and 3 in exactly the hard and full bands; and 9 are detected in all the three standard bands.

\subsubsection{Photometry and Deblending}
\label{sec:phot}
The radius of 90\% encircled-energy fraction contour of the \textit{NuSTAR} point spread function (PSF) is approximately $67.5''$, which is relatively large compared to the average distance among sources. We adopt the similar strategy to that of M15 to choose an aperture size of $30''$ for photometry extraction and assume that the net counts within this aperture are only contaminated by other nearby \textit{NuSTAR}-detected sources within $90''$.\par
For each source in our catalog, the total and background counts are calculated within a circular region of radius $30''$ in the science and background mosaics, respectively, while the net counts are derived by subtracting the background counts from the total counts. Following Gehrels\upcite{Gehrels1986}, we estimate the upper and lower $1\sigma$ confidence limits on the total counts; for those not detected in certain bands, only the upper limits are derived, using:
\begin{equation}
	\centering
\begin{split}
	\sum_{x=0}^{n} \frac{\lambda_{u}^{x} e^{-\lambda_u} }{x!} &= 1-{\rm CL, and} \\
	\sum_{x=0}^{n-1} \frac{\lambda_{l}^{x} e^{-\lambda_l} }{x!} &= {\rm CL,}
\end{split}
\end{equation}
where $\lambda_u$ and $\lambda_l$ represent the upper and lower limits, $n$ is the photon counts, and CL represents the confidence level, respectively.\par
The background count error can be approximated by $\sigma_{\rm Bgd}=\frac{1+\sqrt{factor\!\times\!C_{\rm Bgd}\!+\!3/4}}{factor}$, where $C_{\rm Bgd}$ is the background counts and $factor$ gives the ratio between the total area where the background model is defined and the area for photometry extraction (i.e., $factor=(180''/30'')^2\times 4=144$). Subsequently, the upper and lower limits on the net counts are calculated as $\sigma_{\rm Net,u}=\sqrt{\lambda_u^2+\sigma_{\rm Bgd}^2}$ and $\sigma_{\rm Net,l}=\sqrt{\lambda_l^2+\sigma_{\rm Bgd}^2}$, respectively. \par
To deblend the sources in our catalog, the FOF algorithm is applied again to split them into group sources and isolated sources with a different linking length of $90''$. For the isolated sources, we assume that they cannot be contaminated by any other sources (away beyond $90''$); for the group sources, a system of $n$ linear simultaneous equations is established:
\begin{equation}
	\centering
\begin{split}
	C_T^1 & = N(r_{1,1})C_D^1 + N(r_{1,2})C_D^2 + ... + N(r_{1,n})C_D^n \\
	C_T^2 & = N(r_{2,1})C_D^1 + N(r_{2,2})C_D^2 + ... + N(r_{2,n})C_D^n \\
	...\\
	C_T^n & = N(r_{n,1})C_D^1 + N(r_{n,2})C_D^2 + ... + N(r_{n,n})C_D^n
\end{split}
\end{equation}
where $C_T^n$ is the total net counts of source $n$, $C_D^n$ is the deblended net counts of source $n$, and $N(r_{i,j})$ is the normalized function of the separation between the sources $i$ and $j$ ($r_{i,j}$ represents the separation distance, while $N(0)=1$), in which several simplifications are proposed to avoid the complications of the nonazimuthally symmetric \textit{NuSTAR} PSF.\par
Following the deblending procedure above, we then perform deblending with another aperture of $20''$ radius, and recalculate the $P_{\rm false}$ of each source after deblending. The post-deblending $P_{\rm false}$ values are compared with the $P_{r20}$ thresholds, and 4 of the 58 sources in our catalog become no longer significant. Additionaly, we find 1 source in the area of relatively low exposure ($<40$\,ks, corresponding to $\lesssim10\%$ of the maximum survey exposure). All of these 5 sources are detected in the full band only, and we flag but do not remove them (see Figure~\ref{fig:E-CDF-S_mosaic}).\par
\begin{figure}
\centering
\includegraphics[width=\textwidth]{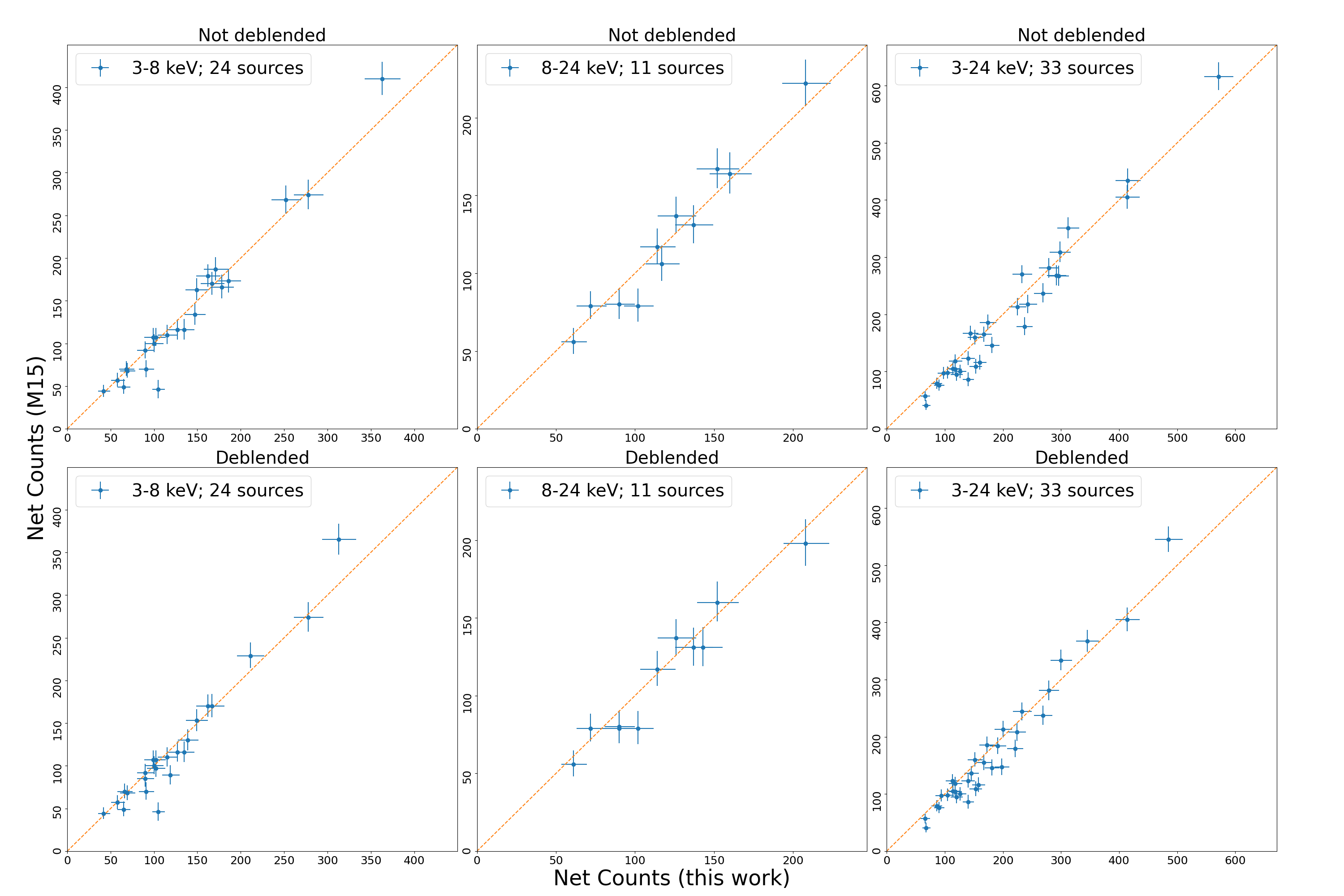}
	\caption{Comparison of net counts obtained by this work and M15 in the E-CDF-S. The diagonal dashed lines indicate the 1:1 relations.}
\label{fig:compare}
\end{figure}

To validate the reliability of our catalog (a total of 58 sources), we match it to the previous \textit{NuSTAR} E-CDF-S catalog (a total of 54 sources) in M15 using a matching radius $r_{\rm m}=30''$ and find a total of 36 counterpart pairs. We compare their net counts in Figure~\ref{fig:compare} and find good consistency within $1\sigma$ errors. We also compare their aperture-corrected fluxes (see Section~\ref{sec:match}) and find good agreement between each other.\par
A significant fraction of the M15 sources are not detected by our work (and vise versa), mainly due to two facts: the detailed cataloging methodologies are different between our work and M15, and those unmatched sources generally have lower net counts such that they could be too faint to be detected by either work.

\subsection{Matching to the \textit{Chandra} E-CDF-S and CDF-S Catalogs}
\label{sec:match}
We first match our catalog to the \textit{Chandra} 250\,ks E-CDF-S catalog (X16) using $r_{\rm m}=30''$, and find 51 of the 58 sources to have at least one \textit{Chandra} counterpart. In these matches, 20, 24, 5, and 2 \textit{NuSTAR} sources have 1, 2, 3, and 4 \textit{Chandra} counterparts, respectively; no \textit{NuSTAR} sources have more than 4 \textit{Chandra} counterparts. For the flagged sources (see Section~\ref{sec:phot}), all of the 4 less-significant sources (XIDs=14, 15, 25, 38; see  Figure~\ref{fig:E-CDF-S_mosaic}) have \textit{Chandra} counterparts, but the source with low exposure time (XID=18) does not. We then match the 7 sources without \textit{Chandra} 250\,ks counterparts to the \textit{Chandra} 7\,Ms CDF-S catalog (L17) using $r_{\rm m}=30''$, only finding one further match (XID=19).\par
We inspect the positions and properties of the 6 unmatched sources, finding that 3 reside in the very edge of the \textit{NuSTAR} E-CDF-S survey mosaic (XIDs=18, 21, and 45; see Figure~\ref{fig:E-CDF-S_mosaic}), which implies uncertainty of their detections. For the 3 remaining sources, one (XID=2) is also near the edge of the mosaic and detected only in the 3--24 keV band, and the other two sources (XIDs=37, 49) reside in the chip-gap areas (see Figure~\ref{fig:E-CDF-S_mosaic}) which might be spurious. Because we aim to find as many sources as possible using our algorithm alone (without manual intervention), these 6 sources are conserved and flagged in our final catalog.\par
To compare the fluxes of these matched sources, the observed deblended fluxes of the \textit{NuSTAR} sources are derived following the same approach as Alexander et al.\upcite{Alexander2013}. For the sources detected both in the soft and hard bands, we calculate their hardness ratios, HR=(H$-$S)/(H+S), using the Bayesian estimation of hardness ratios method\upcite{Park2006}; for other sources, an HR value corresponding to the power-law spectral photon index of $\Gamma=1.8$ is assumed. Using the derived HRs, the same parameters as in M15 are then adopted to convert count rates to observed fluxes (see Section 2.3.3 of M15), which reach a soft-band flux limit of $\approx 10^{-14}$ erg~s$^{-1}$~cm$^{-2}$.\par
\begin{figure}
\centering
\includegraphics[width=0.49\textwidth]{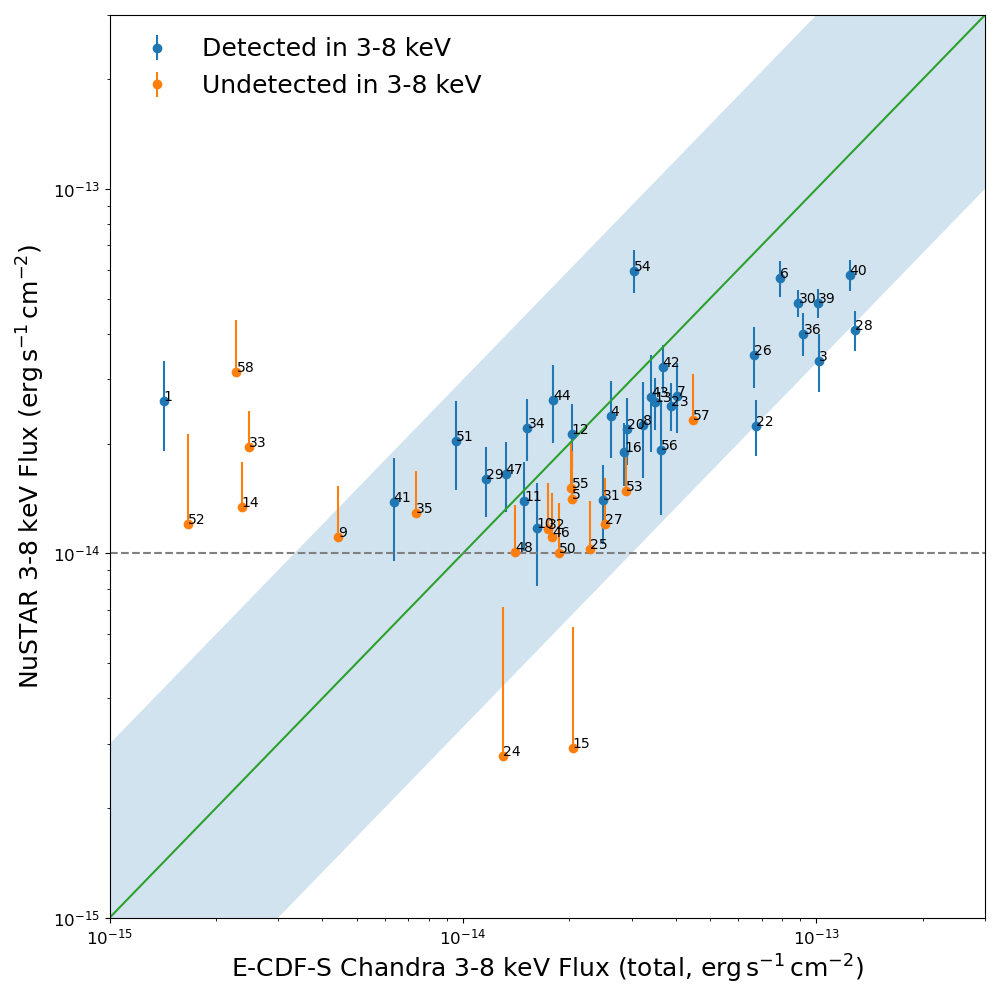}
\includegraphics[width=0.49\textwidth]{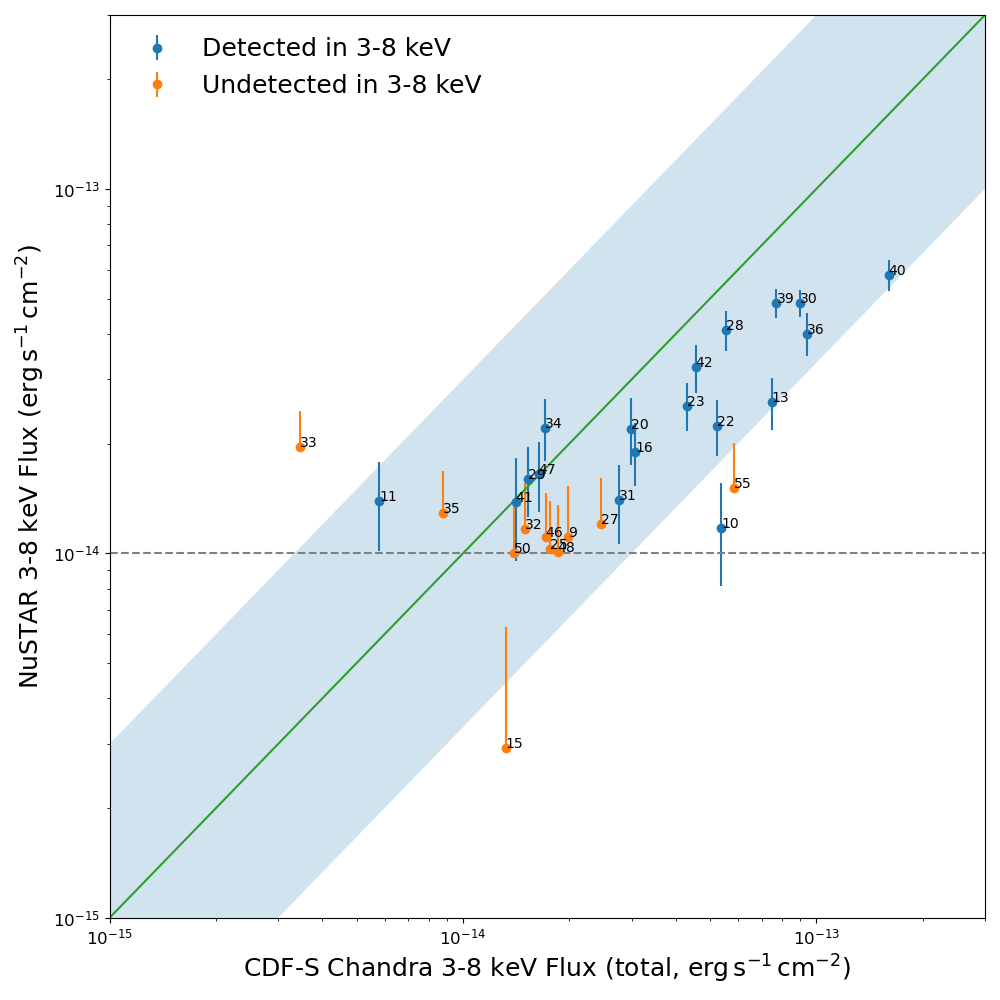}
	\caption{Comparison between the \textit{NuSTAR} E-CDF-S deblended fluxes (this work) and the total \textit{Chandra} E-CDF-S (X16; left panel) and CDF-S (L17; right panel) fluxes in the 3--8 keV band (all fluxes are aperture-corrected). In each panel, the green 1:1 line is centered at the shaded area indicating a factor of $\le 3$ difference from the 1:1 line, and the horizontal dotted line indicates the detection limit.}
\label{fig:flux_cdfs}
\end{figure}

Due to the different observable energy range of \textit{Chandra} (0.5--7 keV), we assume the \textit{Chandra} counterparts having a simple power-law spectrum (i.e., $f(E)\propto E^{-\Gamma}$), then derive the conversion factor between the fluxes in the 2--7 and 3--8 keV bands. The fluxes in the 2--7 keV band and photon indices ($\Gamma$) can be obtained from the \textit{Chandra} catalogs (X16; L17). For the multiple \textit{Chandra} counterparts within $r_{\rm m}=30''$ of the \textit{NuSTAR} sources, we calculate their total 3--8 keV fluxes instead. The comparison between the \textit{NuSTAR} and total \textit{Chandra} 3--8 keV fluxes of the matched sources is shown in Figure~\ref{fig:flux_cdfs}, which indicates general agreement within a factor of 3 for the majority of the sources. However, the \textit{NuSTAR} fluxes appear to be systematically lower than the \textit{Chandra} fluxes, which is mainly due to that the \textit{NuSTAR} measured/assumed photon index may be different from the \textit{Chandra} measured/assumed photon index (in the case of 1-to-1 match) or photon indexes (in the case of 1-to-multiple match).

\section{Production of the \textit{NuSTAR} CDF-N Point-Source Catalog}
\label{sec:CDF-N_cat}
\subsection{Data Reduction and Source Detection}
We collect 12 valid observations from the \textit{NuSTAR} CDF-N survey that cover a sky area of $\approx7'\times10'$, almost each of which has an effective exposure of $\approx$50\,ks. We summarize the information of these observations in Table~\ref{tab:table2}. For data reduction, we apply the exactly same procedures as in Section~\ref{sec:data}, therefore the technical details are introduced briefly in this section.\par

\begin{center}
\begin{ThreePartTable}

\begin{TableNotes}\footnotesize
\item[] \textbf{Notes}: Same as those of Table~\ref{tab:table1}.
\end{TableNotes}

\begin{longtable}{cccccc}
% 首页表头
	\caption{Details of the \textit{NuSTAR} CDF-N Observations}	\label{tab:table2} \\
\toprule[1.5pt]
	Obs.\,ID\tnote{1} & Obs.\,Name\tnote{2} & Obs.\,Date\tnote{3} & RA\tnote{4} & DEC\tnote{5} & $t_{\rm eff}$\tnote{6} \\
\midrule[1pt]
\endfirsthead
% 续页表头
	\caption[]{Details of the \textit{NuSTAR} CDF-N Observations (continued)} \\
\toprule[1.5pt]
	Obs.\,ID\tnote{1} & Obs.\,Name\tnote{2} & Obs.\,Date\tnote{3} & RA\tnote{4} & DEC\tnote{5} & $t_{\rm eff}$\tnote{6} \\
\midrule[1pt]
\endhead
% 首页表尾
\hline
\multicolumn{6}{r}{\textit{Table~\ref{tab:table2} continued}}
\endfoot
% 续页表尾
\bottomrule[1.5pt]
\insertTableNotes
\endlastfoot
	60110003002 & GOODSN\_MOS003 & 2015 Apr 22 & 12.61 & +62.20 & 45.6 \\
	60110002001 & GOODSN\_MOS002 & 2015 Apr 23 & 12.62 & +62.24 & 45.6 \\
	60110001001 & GOODSN\_MOS001 & 2015 Apr 24 & 12.62 & +62.28 & 46.1 \\
	60110001003 & GOODSN\_MOS001 & 2015 Aug 04 & 12.62 & +62.28 & 59.7 \\
	60110002003 & GOODSN\_MOS002 & 2015 Aug 09 & 12.61 & +62.24 & 64.3 \\
	60110003003 & GOODSN\_MOS003 & 2015 Aug 10 & 12.61 & +62.20 & 59.0 \\
	60110001005 & GOODSN\_MOS001 & 2015 Oct 31 & 12.62 & +62.27 & 61.5 \\
	60110002004 & GOODSN\_MOS002 & 2015 Nov 01 & 12.62 & +62.23 & 64.3 \\
	60110003004 & GOODSN\_MOS003 & 2015 Nov 02 & 12.61 & +62.20 & 67.8 \\
	60110001007 & GOODSN\_MOS001 & 2016 Jan 30 & 12.62 & +62.27 & 52.8 \\
	60110002005 & GOODSN\_MOS002 & 2016 Jan 31 & 12.62 & +62.23 & 52.9 \\
	60110003005 & GOODSN\_MOS003 & 2016 Feb 01 & 12.61 & +62.19 & 54.0 \\
\end{longtable}
\end{ThreePartTable}
\end{center}

\begin{figure}
\centering
\includegraphics[width=0.5\textwidth]{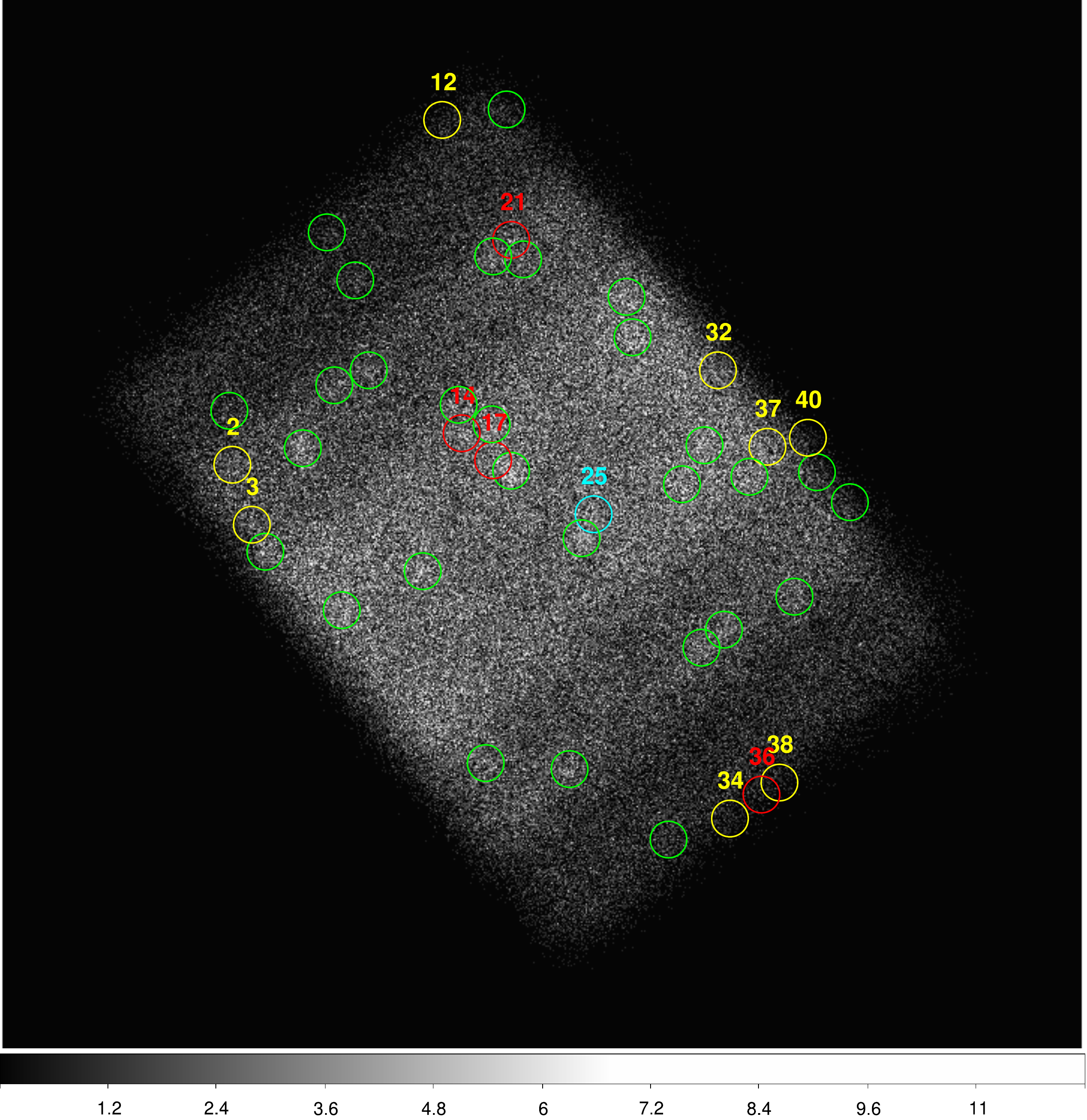}
\caption{Stacked \textit{NuSTAR} CDF-N science mosaic in the 3--24 keV band (cf. Figure~\ref{fig:E-CDF-S_mosaic}), with a total of 42 sources plotted. The green (29/42) and red (4/42; being less significant detections) sources have the \textit{Chandra} 2\,Ms CDF-N counterparts within $r_{\rm m}=30''$. Among the unmatched sources (9/42), 8 yellow sources reside in or near the edges of the mosaic, and 1 cyan source is only detected in the 8--24 keV band.}
\label{fig:CDF-N_mosaic}
\end{figure}

Full-field lightcurves in the entire energy band with a bin size of 20~s are produced to inspect the influence of flaring events, and a threshhold of 1.3 cts\,$\mathrm{s^{-1}}$ is selected to reject the periods with high background flares. Following the same procedures in Section~\ref{sec:mosaic}, the science, exposure, and background mosaics are created from the cleaned event files. The CDF-N reaches a maximum depth of $\sim$ 440 ks corrected for vignetting in the full band, almost doubling that of the E-CDF-S. We present the stacked science mosaic in the full band in Figure~\ref{fig:CDF-N_mosaic}.\par

Again we set $R=0.1$ to obtain the seed list from the $P_{\rm fasle}$ maps, then refilter these seed sources with a strict threshold of $R=0.01$ for the final catalog that contains 42 CDF-N sources, each of which is detected in at least one of the three standard bands. Of the 42 sources, 26, 11, and 31 are detected in the soft, hard, and full bands, respectively; 9, 2, and 12 are detected in the soft, hard, and full bands only, respectively; no source is detected in exactly both the soft and hard bands, 10 in exactly both the soft and full bands, and 2 in exactly both the hard and full bands; 7 are detected in all the three standard bands. After deblending, 6 of the 42 sources are no longer significant, and we still flag and keep them in our final catalog. These results are listed in Table~\ref{tab:seed_CDF-N}.

\begin{table}[th]
\caption{The \textit{NuSTAR} CDF-N Cataloging Process}
\label{tab:seed_CDF-N}
\centering
\resizebox{\textwidth}{!}{
\begin{tabular}{cccccc}\hline\hline
	Seed Catalog		& Source \# 	& Source \# 	& Source \# 	& Source \#	& Source \# \\
	($R=0.1$)		& ($P_{r0})$	& ($P_{r10}$)	& ($P_{r20}$)	& (Total)	& (After Deduplication) \\
	\hline
	3--8 keV 	& 27		& 19		& 28		& 74	& 32		\\
	8--24 keV 	& 4			& 9			& 13		& 26	& 14		\\
	3--24 keV	& 42		& 26		& 30		& 98	& 38		\\
	\hline\hline
	Final Catalog	& Source \# 	& Source \# 	& Source \# 	& Source \# 	& Source \# \\
	  ($R=0.01$ \& $P_{r20}$)     	& (3--8 keV)	& (8--24 keV)	& (3--24 keV)	& (3--8 \& 8--24 keV)	& (Total) \\
	\hline
	Before Deblending	& 26	& 11		& 31		& 7		& 42		\\
	After Deblending & 22	& 11		& 27		& 6		& 36		\\
	\hline
\end{tabular}
}
\end{table}

\subsection{Matching to the \textit{Chandra} CDF-N Catalog}
We match our catalog to the \textit{Chandra} 2\,Ms CDF-N catalog (X16) using $r_{\rm m}=30''$, and find 33 of the 42 sources to have at least one \textit{Chandra} counterpart. In these matches, 14, 11, 4, and 3 \textit{NuSTAR} sources have 1, 2, 3, and 4 \textit{Chandra} counterparts, respectively, and 1 \textit{NuSTAR} source has more than 4 \textit{Chandra} counterparts.\par
For the \textit{NuSTAR} sources without \textit{Chandra} counterparts, we inspect their positions and properties and find that almost all of them (8/9; XIDs=2, 3, 12, 32, 34, 37, 38, 40) reside in or near the edges of the \textit{NuSTAR} CDF-N field and the remaining one (XID=25) is only detected in the hard band which might be too ``hard" to be detected by \textit{Chandra} (see Figure~\ref{fig:CDF-N_mosaic}).\par
\begin{figure}
\centering
\includegraphics[width=0.5\textwidth]{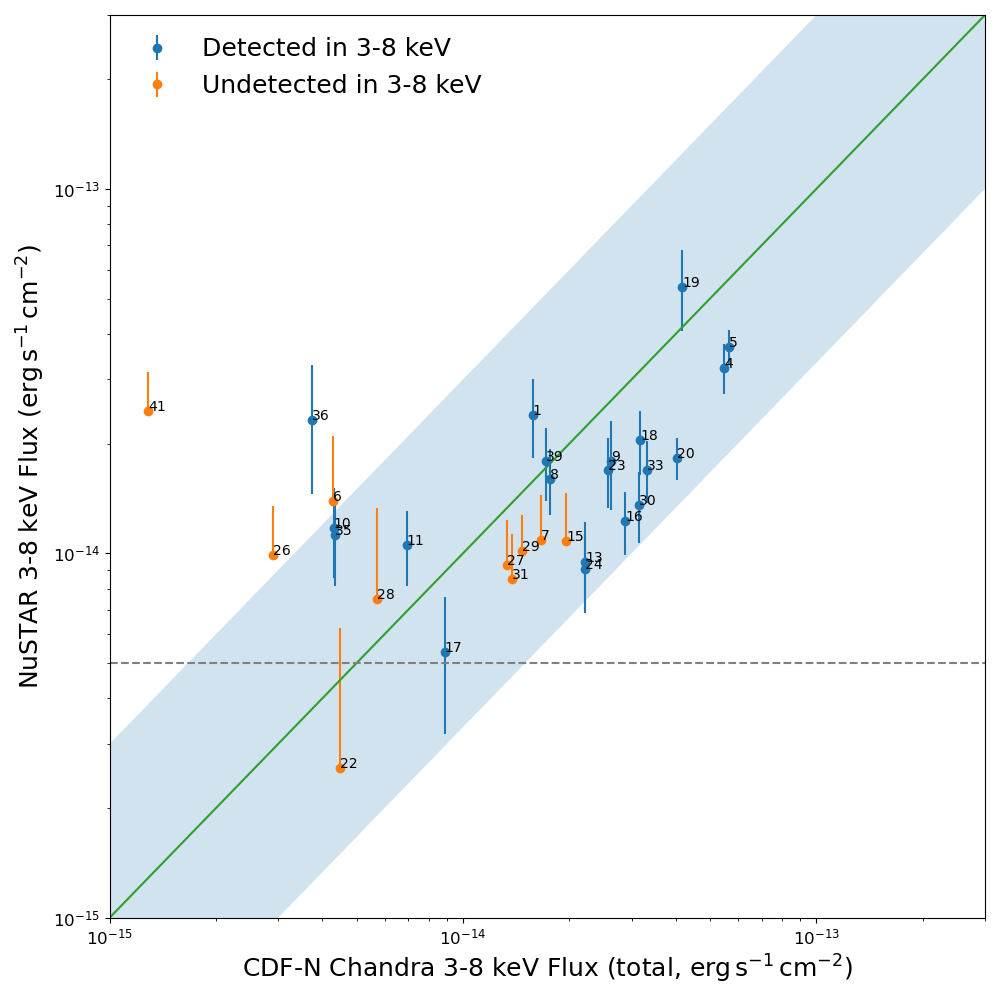}
	\caption{Same as Figure \ref{fig:flux_cdfs}, but for the comparison between the \textit{NuSTAR} CDF-N deblended fluxes (this work) and the total \textit{Chandra} CDF-N fluxes (X16) in the 3--8 keV band (all fluxes are aperture-corrected)}.
\label{fig:flux_CDF-N}
\end{figure}
\begin{figure}
\centering
\includegraphics[width=\textwidth]{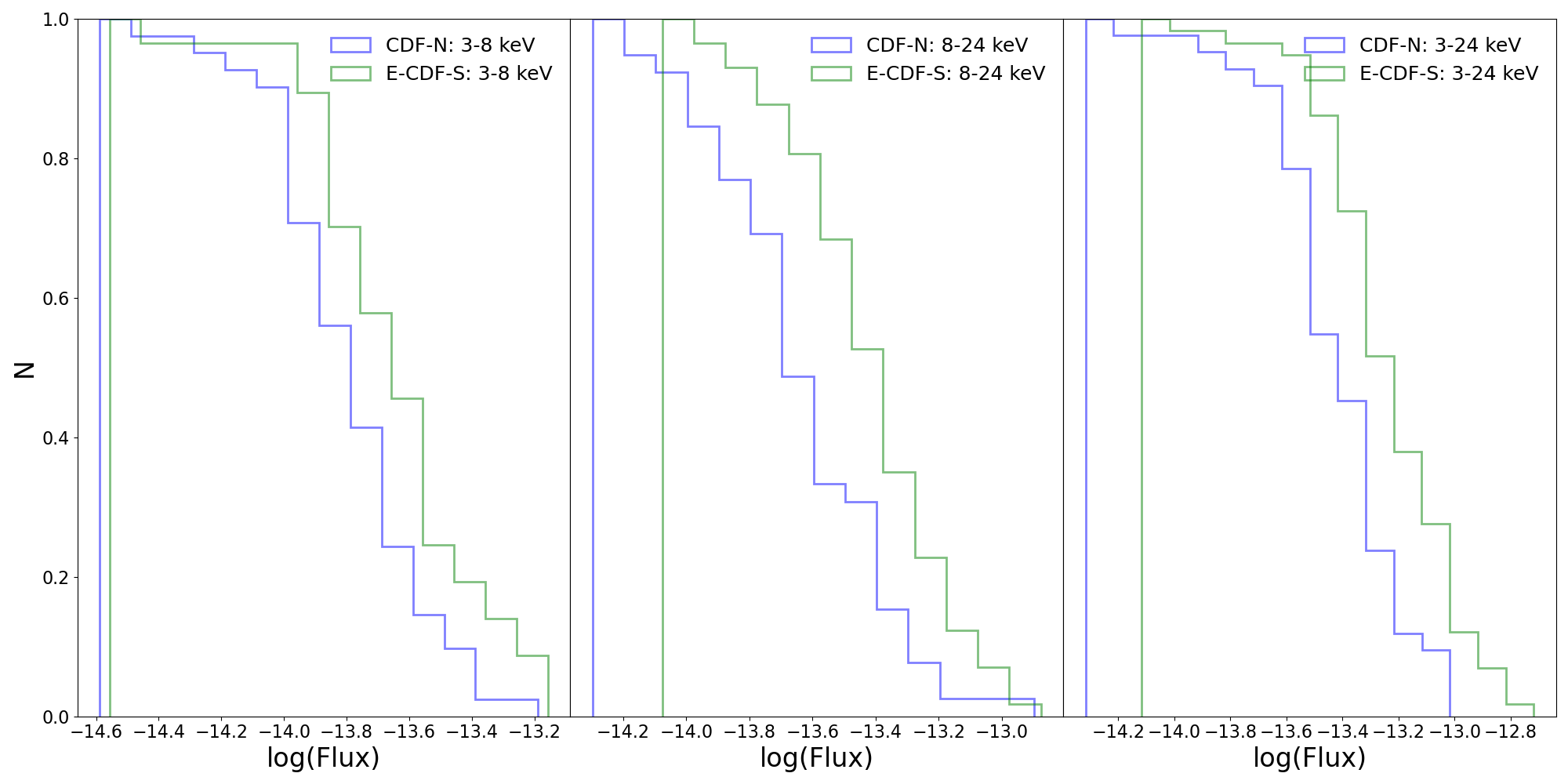}
\caption{Normalized distributions of deblended \textit{NuSTAR} fluxes in the CDF-N and E-CDF-S.}
\label{fig:flux_cmp}
\end{figure}

We compare the fluxes of these matched sources in Figure~\ref{fig:flux_CDF-N}, which also indicates general agreement within a factor of 3 for the majority of the sources.
The normalized flux histograms of the \textit{NuSTAR} CDF-N and E-CDF-S sources are compared in Figure~\ref{fig:flux_cmp}: CDF-N sources generally have lower fluxes than  E-CDF-S sources, being consistent with the fact that the average \textit{NuSTAR} CDF-N exposure depth (reaching a soft-band flux limit of $\approx 5\times 10^{-15}$ erg~s$^{-1}$~cm$^{-2}$) is almost twice that of the E-CDF-S.\par

\section{Summary}
\label{sec:summary}
In this work, we collect the original observations from the \textit{NuSTAR} E-CDF-S and CDF-N surveys and produce cleaned event files. Simulated background mosaics are generated using the IDL software \textit{nuskybgd}, and then processed along with science mosaics to produce $P_{\rm false}$ maps for source detection.\par
For the \textit{NuSTAR} E-CDF-S survey, the main results are as follows:
\begin{enumerate}
\item
The E-CDF-S catalog consists of 58 sources that are detected using our algorithm without manual intervention, with 4 of them being not significant after deblending.
\item
We compare our catalog with the previous \textit{NuSTAR} E-CDF-S catalog (M15) using $r_{\rm m}=30''$ and find a total of 36 matches, the net counts of which agree well within $1\sigma$ errors.
\item
We compare our catalog with the \textit{Chandra} E-CDF-S and CDF-S catalogs (X16 and L17), and find a total of 51 matches, the fluxes of which agree well above the detection limit. All of the 4 sources being not significant after deblending have counterparts in the \textit{Chandra} catalogs; and the 7 unmatched sources are flagged as being spurious but still conserved in the catalog.
\end{enumerate}

For the \textit{NuSTAR} CDF-N survey, the main results are as follows:
\begin{enumerate}
\item
The CDF-N catalog, produced for the first time by this work, consists of 42 sources that are detected using our algorithm without manual intervention, with 6 of them being not significant after deblending.
\item
We compare our catalog with the \textit{Chandra} CDF-N catalog (X16), and find a total of 33 matches, the fluxes of which agree well above the detection limit. We flag the 9 unmatched sources as being spurious but conserve them in the catalog.
\item
The flux limits are significantly lower in the \textit{NuSTAR} CDF-N field (having deeper exposures) than that in the \textit{NuSTAR} E-CDF-S field.
\end{enumerate}

Our source-detection method provides a systematic approach for source cataloging in other \textit{NuSTAR} extragalactic surveys.
We make our \textit{NuSTAR} E-CDF-S and CDF-N source catalogs publicly available (see Appendix~\ref{app:cat-overview} for catalog description), which provide a uniform platform that facilitates further studies involving these two fields.

\section*{Acknowledgements}
We thank the reviewers for helpful comments that improved this work. T.Y.Z. and Y.Q.X. acknowledge support from NSFC grants (12025303 and 11890693), National Key R\&D Program of China No. 2022YFF0503401, the K.C. Wong Education Foundation, and the science research grants from the China Manned Space Project with NO. CMS-CSST-2021-A06.

\section*{Conflict of interest}
The authors declare that they have no conflict of interest.

\section*{Biographies}
\textbf{Tianyi Zhang} is a former graduate student under the supervision of Professor Yongquan Xue at University of Science and Technology of China. His research is focused on active galactic nuclei.\\
\textbf{Yongquan Xue} received his Ph.D. degree from Purdue University. He is currently a professor at University of Science and Technology of China. His research interests mainly focus on active galactic nuclei and high-energy astrophysics.

\appendix
\section{Catalog Description}
\label{app:cat-overview}

\begin{table}[th]
\caption{Overview of Columns in the \textit{NuSTAR} E-CDF-S and CDF-N Catalogs}
\label{tab:cat-overview}
\centering
\resizebox{\textwidth}{!}{
\begin{tabular}{l@{\hspace{1.5cm}}l}\hline\hline
	Column(s) & Description \\
	\hline
	1		&	Source sequence number (i.e., XID) in this work	\\
	2, 3	&	J2000.0 RA and DEC of the \textit{NuSTAR} source		\\
	4		&	Flag of whether the exposure time of the source is above the threshold	\\
	5--7	&	Flags indicating in which of the three standard bands the source is detected	\\
	8--10	&	Flags indicating in which of the three standard bands the source is detected after deblending	\\
	11--13	&	$P_{\rm false}$ values for the three standard bands	\\
	14--16	&	$P_{\rm false}$ values for the three standard bands after deblending	\\
	17		&	Flag of whether the source remains significant after deblending	\\
	18--26 	&	Net counts and associated errors in the three standard bands after deblending	\\
	27--29	& 	Effective exposure times in the three standard bands	\\
	30--38	&	Fluxes and associated errors in the three standard bands after deblending	\\
	39--41	&	Hardness ratio and associated errors	\\
	42--45 	& 	Catalog name, sequence number, RA, and DEC of the \textit{Chandra} counterpart	\\
	46		& 	Separation between the \textit{NuSTAR} position and the \textit{Chandra} counterpart	\\
	47		&	Flux of the \textit{Chandra} counterpart in the 3--8 keV band	\\
	48 		& 	Total fluxes of the \textit{Chandra} counterparts within $30''$ in the 3--8 keV band	\\
	49--50	&	Photometric redshift and spectroscopic redshift of the \textit{Chandra} counterpart	\\
	51		&	Adopted redshift of the \textit{NuSTAR} source	\\
	\hline
\end{tabular}
}
\end{table}

The \textit{NuSTAR} E-CDF-S and CDF-N source catalogs have the same 51 columns that are summarized in Table~\ref{tab:cat-overview}, with the details described below.
\begin{enumerate}
	\item Column 1 gives the source sequence number (i.e., XID). We list the sources in the order of decreasing right ascension.
	\item Columns 2 and 3 give the right ascension and declination of the \textit{NuSTAR} source, respectively.
	\item Column 4 gives the flag whether the exposure time of the source is above the threshold of $\approx 40$\,ks.
	\item Columns 5--7 give the flags indicating in which of the three standard bands (3--8, 8--24 and 3--24 keV bands) the source is detected.
	\item Columns 8--10 give the flags indicating in which of the three standard bands the source is detected after deblending.
	\item Columns 11--13 give the $P_{\rm false}$ values for the three standard bands.
	\item Columns 14--16 give the $P_{\rm false}$ values for the three standard bands after deblending.
	\item Column 17 gives the flag whether the source remains significant after deblending.
	\item Columns 18--20 give the post-deblending net counts with the corresponding lower and upper errors in the 3--8 keV band.
	\item Columns 21--23 give the post-deblending net counts with the corresponding lower and upper errors in the 8--24 keV band.
	\item Columns 24--26 give the post-deblending net counts with the corresponding lower and upper errors in the 3--24 keV band.
	\item Columns 27--29 give the effective exposure times derived from the exposure maps for the three standard bands.
	\item Columns 30--32 give the post-deblending flux with the corresponding lower and upper errors in units of $10^{-14}\,\mathrm{erg\,s^{-1}\,cm^{-2}}$ in the 3--8 keV band (hereafter the same units).
	\item Columns 33--35 give the post-deblending flux with the corresponding lower and upper errors in the 8--24 keV band.
	\item Columns 36--38 give the post-deblending flux with the corresponding lower and upper errors in the 3--24 keV band.
	\item Columns 39--41 give the hardness ratio with the corresponding lower and upper errors.
	\item Columns 42--46 give the \textit{Chandra} counterpart with the corresponding catalog name, source sequence number, right ascension and declination, and  distance (in units of arcsecond) from the \textit{NuSTAR} source.
	\item Column 47 gives the flux of the \textit{Chandra} counterpart in the 3--8 keV band.
	\item Column 48 gives the total flux of the \textit{Chandra} counterparts found within $30''$ in the 3--8 keV band.
	\item Columns 49 and 50 give the photometric redshift and spectroscopic redshift of the \textit{Chandra} counterpart.
	\item Column 51 gives the adopted redshift of the source, with the spectroscopic redshift preferred if available.
\end{enumerate}

\bibliographystyle{custom-unsrt}
\bibliography{bibtex}

\begin{thebibliography}{10}

\bibitem{Brandt2005}
W.~N. Brandt and G.~Hasinger.
\newblock {Deep extragalactic X-ray surveys}.
\newblock \textbf{2005}, {\em Annual Review of Astronomy and Astrophysics},
  \textit{43}: 827--859.

\bibitem{Brandt2015}
W.~N. Brandt and D.~M. Alexander.
\newblock Cosmic X-ray surveys of distant active galaxies.
\newblock \textbf{2015}, {\em The Astronomy and Astrophysics Review},
  \textit{23} (1): 1.

\bibitem{Xue:2017rpk}
Y.~Q. Xue.
\newblock {The Chandra Deep Fields: Lifting the Veil on Distant Active Galactic
  Nuclei and X-Ray Emitting Galaxies}.
\newblock \textbf{2017}, {\em New Astron. Rev.}, \textit{79}: 59--84.

\bibitem{YN2014}
Feng {Yuan} and Ramesh {Narayan}.
\newblock {Hot Accretion Flows Around Black Holes}.
\newblock \textbf{2014}, {\em Annual Review of Astronomy and Astrophysics},
  \textit{52}: 529--588.

\bibitem{Li2019}
Junyao {Li}, Yongquan {Xue}, Mouyuan {Sun}, and et~al.
\newblock {Piercing through Highly Obscured and Compton-thick AGNs in the
  Chandra Deep Fields. I. X-Ray Spectral and Long-term Variability Analyses}.
\newblock \textbf{2019}, {\em The Astrophysical Journal}, \textit{877} (1): 5.

\bibitem{Li2020}
Junyao {Li}, Yongquan {Xue}, Mouyuan {Sun}, and et~al.
\newblock {Piercing through Highly Obscured and Compton-thick AGNs in the
  Chandra Deep Fields. II. Are Highly Obscured AGNs the Missing Link in the
  Merger-triggered AGN-Galaxy Coevolution Models?}
\newblock \textbf{2020}, {\em The Astrophysical Journal}, \textit{903} (1): 49.

\bibitem{Hickox2006}
Ryan~C. Hickox and Maxim Markevitch.
\newblock Absolute Measurement of the Unresolved Cosmic X-Ray Background in the
  0.5--8 {keV} Band with \textit{Chandra}.
\newblock \textbf{2006}, {\em The Astrophysical Journal}, \textit{645} (1):
  95--114.

\bibitem{Xue2011a}
Y.~Q. Xue, B.~Luo, W.~N. Brandt, and et~al.
\newblock {THE CHANDRA DEEP FIELD-SOUTH SURVEY: 4 Ms SOURCE CATALOGS}.
\newblock \textbf{2011}, {\em The Astrophysical Journal Supplement},
  \textit{195} (1): 10.

\bibitem{Xue2012}
Y.~Q. {Xue}, S.~X. {Wang}, W.~N. {Brandt}, and et~al.
\newblock {Tracking down the Source Population Responsible for the Unresolved
  Cosmic 6-8 keV Background}.
\newblock \textbf{2012}, {\em The Astrophysical Journal}, \textit{758} (2):
  129.

\bibitem{Lehmer2012}
B.~D. Lehmer, Y.~Q. Xue, W.~N. Brandt, and et~al.
\newblock {THE 4 Ms CHANDRA DEEP FIELD-SOUTH NUMBER COUNTS APPORTIONED BY
  SOURCE CLASS: PERVASIVE ACTIVE GALACTIC NUCLEI AND THE ASCENT OF NORMAL
  GALAXIES}.
\newblock \textbf{2012}, {\em The Astrophysical Journal}, \textit{752} (1): 46.

\bibitem{Luo2017}
B.~Luo, W.~N. Brandt, Y.~Q. Xue, and et~al.
\newblock {THE \textit{CHANDRA} DEEP FIELD-SOUTH SURVEY: 7 MS SOURCE CATALOGS}.
\newblock \textbf{2017}, {\em The Astrophysical Journal Supplement},
  \textit{228} (1): 2.

\bibitem{Harrison2016}
F.~A. Harrison, J.~Aird, F.~Civano, and et~al.
\newblock { THE NuSTAR EXTRAGALACTIC SURVEYS: THE NUMBER COUNTS OF ACTIVE
  GALACTIC NUCLEI AND THE RESOLVED FRACTION OF THE COSMIC X-RAY BACKGROUND }.
\newblock \textbf{2016}, {\em The Astrophysical Journal}, \textit{831} (2):
  185.

\bibitem{Harrison2013a}
Fiona~a. Harrison, William~W. Craig, Finn~E. Christensen, and et~al.
\newblock {The Nuclear Spectroscopic Telescope Array (\textit{NuSTAR})
  Mission}.
\newblock \textbf{2013}, {\em The Astrophysical Journal}, \textit{770} (1): 20.

\bibitem{Xue2016a}
Y.~Q. Xue, B.~Luo, W.~N. Brandt, and et~al.
\newblock { THE 2 Ms CHANDRA DEEP FIELD-NORTH SURVEY AND THE 250 Ks EXTENDED
  CHANDRA DEEP FIELD-SOUTH SURVEY: IMPROVED POINT-SOURCE CATALOGS }.
\newblock \textbf{2016}, {\em The Astrophysical Journal Supplement},
  \textit{224} (2): 15.

\bibitem{Mullaney2015a}
J.~R. Mullaney, A.~Del-Moro, J.~Aird, and et~al.
\newblock {THE \textit{NuSTAR} EXTRAGALACTIC SURVEYS: INITIAL RESULTS AND
  CATALOG FROM THE EXTENDED \textit{CHANDRA} DEEP FIELD SOUTH}.
\newblock \textbf{2015}, {\em The Astrophysical Journal}, \textit{808} (2):
  184.

\bibitem{Alexander2013}
D.~M. Alexander, D.~Stern, A.~{Del Moro}, and et~al.
\newblock {THE \textit{NuSTAR} EXTRAGALACTIC SURVEY: A FIRST SENSITIVE LOOK AT
  THE HIGH-ENERGY COSMIC X-RAY BACKGROUND POPULATION}.
\newblock \textbf{2013}, {\em The Astrophysical Journal}, \textit{773} (2):
  125.

\bibitem{Wik2014a}
Daniel~R. Wik, A.~Hornstrup, S.~Molendi, and et~al.
\newblock {NuSTAR OBSERVATIONS OF THE BULLET CLUSTER: CONSTRAINTS ON INVERSE
  COMPTON EMISSION}.
\newblock \textbf{2014}, {\em The Astrophysical Journal}, \textit{792} (1): 48.

\bibitem{Freeman2002}
P.~E. Freeman, V.~Kashyap, R.~Rosner, and et~al.
\newblock A Wavelet-Based Algorithm for the Spatial Analysis of Poisson Data.
\newblock \textbf{2002}, {\em The Astrophysical Journal Supplement Series},
  \textit{138} (1): 185--218.

\bibitem{AE2010}
Patrick~S. {Broos}, Leisa~K. {Townsley}, Eric~D. {Feigelson}, and et~al.
\newblock {Innovations in the Analysis of Chandra-ACIS Observations}.
\newblock \textbf{2010}, {\em The Astrophysical Journal}, \textit{714} (2):
  1582--1605.

\bibitem{Masini2018}
A.~Masini, F.~Civano, A.~Comastri, and et~al.
\newblock The \textit{NuSTAR} Extragalactic Surveys: Source Catalog and the
  Compton-thick Fraction in the {UDS} Field.
\newblock \textbf{2018}, {\em The Astrophysical Journal Supplement},
  \textit{235} (1): 17.

\bibitem{Georgakakis2008}
A.~Georgakakis, K.~Nandra, E.~S. Laird, and et~al.
\newblock {A new method for determining the sensitivity of X-ray imaging
  observations and the X-ray number counts}.
\newblock \textbf{2008}, {\em Monthly Notices of the Royal Astronomical
  Society}, \textit{388} (3): 1205--1213.

\bibitem{Bertin1996}
E.~Bertin and S.~Arnouts.
\newblock {SExtractor: Software for source extraction}.
\newblock \textbf{1996}, {\em Astronomy and Astrophysics Supplement Series},
  \textit{117} (2): 393--404.

\bibitem{Feng2017}
Y.~{Feng} and C.~{Modi}.
\newblock {A fast algorithm for identifying friends-of-friends halos}.
\newblock \textbf{2017}, {\em Astronomy and Computing}, \textit{20}: 44--51.

\bibitem{Civano2015a}
F.~Civano, R.~C. Hickox, S.~Puccetti, and et~al.
\newblock THE \textit{NuSTAR} EXTRAGALACTIC SURVEYS: OVERVIEW AND CATALOG FROM
  THE COSMOS FIELD.
\newblock \textbf{2015}, {\em The Astrophysical Journal}, \textit{808} (2):
  185.

\bibitem{Gehrels1986}
Neil Gehrels.
\newblock {Errors Calculation}.
\newblock \textbf{1986}, {\em The Astrophysical Journal}, \textit{303}:
  336--346.

\bibitem{Park2006}
Taeyoung {Park}, Vinay~L. {Kashyap}, Aneta {Siemiginowska}, and et~al.
\newblock {Bayesian Estimation of Hardness Ratios: Modeling and Computations}.
\newblock \textbf{2006}, {\em The Astrophysical Journal}, \textit{652} (1):
  610--628.

\end{thebibliography}

\end{document}